\documentclass[traditabstract]{aa} 
\usepackage{natbib}
\usepackage{longtable}
\usepackage{lscape}
\usepackage{graphicx}
\usepackage{txfonts}
\usepackage{multirow}
\begin{document}
   \title{Optical and infrared properties of active galactic nuclei in the Lockman Hole}

   \author{E. Rovilos\inst{1}\fnmsep\thanks{send off-print requests to erovilos@mpe.mpg.de}
           \and
           S. Fotopoulou\inst{2,3}
           \and
           M. Salvato\inst{2,4}
           \and
           V. Burwitz\inst{1}
           \and
           E. Egami\inst{5}
           \and
           G. Hasinger\inst{2}
           \and
           G. Szokoly\inst{1,6}
          }
   \institute{Max Planck Institut f\"{u}r extraterrestrische Physik,
              Giessenbachstra\ss e, 85748 Garching, Germany
              \and
              Max Planck Institut f\"{u}r Plasmaphysik, Boltzmannstra\ss e 2, 85748, Garching, Germany
              \and
              Technische Universit\"{a}t M\"{u}nchen, Fakult\"{a}t f\"{u}r
              Physik, James-Frank-Stra\ss e, 85748 Garching, Germany
              \and
              Excellence Cluster Universe, TUM, Boltzmannstra\ss e 2, 85748, Garching, Germany
              \and
              Steward Observatory, University of Arizona, 933 North Cherry Avenue, Tucson,
              AZ 85721, USA
              \and
              Institute of Physics, E\"{o}tv\"{o}s University, P\'{a}zm\'{a}ny P. s. 1/A,
              1117 Budapest, Hungary
             }

   \date{Draft version of \today}

 \abstract{We present the observed-frame optical, near- and mid-infrared
           properties of X-ray selected AGN in the Lockman Hole. Using a
           likelihood ratio method on optical, near-infrared or mid-infrared
           catalogues, we assigned counterparts to 401 out of the 409  X-ray
           sources of the XMM-Newton catalogue. Accurate photometry was
           collected for all the sources from $U$ to $\rm 24\,\mu m$.
           We used X-ray and optical criteria to remove any normal galaxies,
           galactic stars, or X-ray clusters among them and studied the
           multi-wavelength properties of the remaining 377 AGN. We used a
           mid-IR colour-colour selection to understand the AGN contribution to
           the optical and infrared emission. Using this selection, we
           identified different behaviours of AGN-dominated and host-dominated
           sources in X-ray-optical-infrared colour-colour diagrams. More
           specifically, the AGN dominated sources show a clear trend in the
           $f_{\rm x}/f_{R_C}$ vs. $R_C-K$ and $f_{24\,\mu{\rm m}}/f_{R_C}$ vs.
           $R_C-K$ diagrams, while the hosts follow the behaviour of non X-ray
           detected galaxies. In the optical-near-infrared colour-magnitude
           diagram we see that the known trend of redder objects being more
           obscured in X-rays is stronger for AGN-dominated than for
           host-dominated systems. This is an indication that the trend is more
           related to the AGN contaminating the overall colours than any
           evolutionary effects. Finally, we find that a significant fraction
           ($\sim30\%$) of the reddest AGN are not obscured in X-rays.}
   {}{}{}{}

   \keywords{Galaxies: active -- Galaxies: Seyfert -- Galaxies: statistics -- X-rays: galaxies -- Infrared: galaxies}

   \maketitle
%

\section{Introduction}

Active galactic nuclei (AGN) are among the most energetic phenomena in the
universe being responsible for a significant fraction ($\sim15\%$) of its total
luminosity \citep*{Elvis2002}. The energy output of a single AGN can be as high
as $10^{14}-10^{15}L_{\odot}\simeq10^{47}\,{\rm erg\,s^{-1}}$
\citep*[e.g.][]{Hopkins2007}. The peak of their energy distribution is at high
energies (UV / soft X-rays), hence X-ray surveys are the most efficient and the
most widely used method to detect AGN \citep{Brandt2005}, because of their
small contamination by non-AGN sources.

However, in order to have a complete picture of the properties of AGN, one
needs to take into consideration their full multi-wavelength energy
distribution. While the ionising radiation of AGN can be directly detected in
X-ray and ultra-violet wavelengths \citep{Strateva2005}, there are many cases
where the line of sight is obscured by circumnuclear material. This may absorb
the UV and soft X-ray photons in a quantity depending on its column density.
The fraction of obscured AGN is higher at non-local redshifts
\citep{LaFranca2005,Treister2006,Hasinger2008}, which is reflected on the
characteristic shape of the X-ray background \citep*{Ueda2003,Gilli2007}. The
obscured ionising photons are then re-emitted in infrared wavelengths, where
the resulting SED will show a characteristic power-law spectrum
\citep{Neugebauer1979}. The detection of AGN through this characteristic
feature is a powerful and widely used method of detecting AGN
\citep{Lacy2004,Stern2005,AlonsoHerrero2006,Donley2007}, while adding
information from the optical bands can refine the selection
\citep{Richards2006}. The infrared selection of AGN is particularly important
for the detection of X-ray obscured AGN which soft X-ray and UV light are
absorbed by circumnuclear dust \citep[e.g.][]{Lacy2007,Eckart2010}.

Moreover, the evolution of the AGN cannot be thoroughly studied
without taking into account the properties of the host galaxy, which emits
predominantly in optical and infrared wavelengths. There is an observed
correlation between host and AGN evolution,
\citep[the M-sigma and M-bulge relations;][]{Kormendy1983,Magorrian1998,Ferrarese2000,Gebhardt2000,Tremaine2002,Haring2004,Gultekin2009},
and its origin and evolution with time is still under debate
\citep{Woo2008,Merloni2010}. For the study of the properties of both the AGN
and the host galaxy it is necessary that one separates the observational
characteristics of these two components. In cases of objects at non-local
redshifts, where their angular sizes are usually too small to be resolved by
ground-based optical telescopes without the use of adaptive optics, the only
way to do it is through detecting AGN and host galaxy characteristics in their
broad-band spectra \citep[see][]{Merloni2010}, therefore a multi-wavelength
approach is essential.

In this paper, we present observed-frame multi-wavelength results from the
Lockman Hole survey, which is one of the deepest in X-ray wavelengths
\citep{Brunner2008} and also has broad coverage in optical, near-, and
mid-infrared wavelengths with the most sensitive ground-based (LBT, Subaru, and
UKIRT) and space ({\it Spitzer}) telescopes. In Section\,\ref{datasec} we
describe the multi-wavelength data used and in Section\,\ref{construction} we
use the likelihood ratio method to find counterparts for the X-ray sources and
construct the multi-wavelength catalogue. In Section\,\ref{AGN_select} we
identify any normal galaxies, stars or X-ray clusters, which we remove from our
sample, and in Section\,\ref{core_host} we use the mid-infrared colours to
assess the host galaxy contribution in the optical-infrared flux. Finally, in
Section\,\ref{discussion} we discuss the different behaviours of the hosts and
the AGN, and summarise our results in Section\,\ref{conclusions}.

\section{Data}
\label{datasec}

\subsection{X-rays}

The X-ray observations of the Lockman Hole took place between April 2000 and
December 2002 with XMM-Newton. The final catalogue of the inner 15\arcmin\
of the survey contains 409 sources with flux limits of
$1.9\times10^{-16}{\rm erg\,cm^{-2}s^{-1}}$,
$9\times10^{-16}{\rm erg\,cm^{-2}s^{-1}}$, and
$1.8\times10^{-15}{\rm erg\,cm^{-2}s^{-1}}$ in the 0.5-2.0\,keV, 2.0-10.0\,keV,
and 5.0-10.0\,keV bands
respectively \citep{Brunner2008}. Low resolution X-ray spectra have been
extracted for the brightest 143 sources \citep{Mainieri2002,Mateos2005}, giving
a mixed bag of obscured and unobscured sources, in terms of $N_{\rm H}$. As
these spectra and consequently the hydrogen column density information are
available for only a fraction of the sources, in this paper we will use the
X-ray hardness ratio between the (0.5-2.0)\,keV and (2.0-4.5)\,keV bands as an
indication of the X-ray absorption. \citet{Mainieri2002} use the X-ray and
optical spectra of the X-ray brightest AGN and find that optically type-2 AGN
tend to have large hydrogen column densities
($\log N_{\rm H}>21.5\,{\rm cm}^{-2}$). The dividing line in hardness ratio
between obscured and unobscured sources is at $HR=-0.4$
\citep[Fig.\,3 in][]{Mainieri2002,Hasinger2001}, and this is the value we will
use (see also similar results in \citealt{DellaCeca2004} and simulations in
\citealt{Dwelly2005}).

\subsection{Optical}

The optical observations of the Lockman Hole were conducted with the
Large Binocular Telescope ($U$-$B$-$V$ bands) and the Subaru Telescope
($R_C$-$I_C$-$z'$ bands).
The LBT observations were taken from February 2007 to March 2009. The data
and the analysis technique are described in \citet{Rovilos2009}.
The $R_C$, $I_C$, and $z'$ bands have been observed with the Suprime-Cam of the
Subaru telescope between November 2001 and April 2002 \citep[see][]{Barris2004}.
The data have been analysed using standard techniques and the final images
cover an area of 0.5\,deg$^2$ covering the entire LBT and XMM area. The PSF
FWHM of the final images is in the order of 0.85\,arcsec.

In order to extract the optical magnitudes we first register all the optical
images to a common frame, and as such we use the SWIRE infrared extragalactic
survey \citep{Lonsdale2003}, which has an astrometric accuracy of 0.2\arcsec\
with respect to the 2MASS survey. The typical rms of the positional differences
between the final positions of our sources and the SWIRE sources is one pixel,
which corresponds to 0.2\arcsec, or one quarter to one fifth of the FWHM of the
PSF of the optical images.

We extract the sources of the optical images using sextractor \citep{Bertin1996}
in dual mode with parameters similar to what described in \citet{Rovilos2009}.
As a detection image we use the $R_C$ image, which is the best in terms of
seeing (0.9\arcsec) and depth, and measure the fluxes of the sources in
3\arcsec\ diameter apertures. The largest PSF is that of the $U$-band image
(FWHM=1.06\arcsec) which is 17\% larger than the best PSF. As the seeing is
relatively good, in 3\arcsec\ aperture the total flux is included in all bands
for point sources. Nevertheless, we make simulations to find any residual
aperture corrections needed and fine-tune the final
zero-points using colour-colour track of stars. Details on the relative
photometry will be given in a subsequent paper (Fotopoulou et al. in
preparation).

\subsection{Near- and mid-infrared}

The region of the Lockman Hole in study has been targeted in both the near
infrared by the UKIRT and the mid infrared by {\it Spitzer}-IRAC and
{\it Spitzer}-MIPS. The UKIRT observations are part of the UKIDSS survey
\citep{Lawrence2007}. We use the Data Release 5 (DR5) of the Deep Extragalactic
Survey (DXS), which includes $J$ and $K$-band images and source catalogues
of the Lockman Hole to a limiting magnitude of $K=21$(Vega) in the deepest part
of the survey.

IRAC on board {\it Spitzer} has observed the Lockman Hole in $3.6\,\mu{\rm m}$,
$4.5\,\mu{\rm m}$, $5.8\,\mu{\rm m}$, and $8.0\,\mu{\rm m}$ in April 2004 with
a total integration time of 500\,s per band
\citep[see details in][]{PerezGonzalez2008}. We have used sextractor in dual
mode, using the $3.6\,\mu{\rm m}$ image to detect sources and measured their
fluxes in all four bands. We have extracted aperture magnitudes with an
aperture diameter of 3.8\arcsec in all four bands and used standard aperture
corrections \citep{Surace2005} to derive the total magnitudes. The detection
limit at $3.6\,\mu{\rm m}$ is 24.5\,mag(AB). This method may cause an
underestimation of the fluxes of extended sources, however our analysis shows
that the number of affected sources is less than 20\% of the $3.6\,\mu{\rm m}$
detections. Moreover, this aperture effect is less severe in the IRAC colours
used in this paper due to the similar PSF sizes of the IRAC channels 1-2 and
3-4.

The MIPS observations were conducted on November 2003, with a total integration
time of 300\,s\,pixel$^{-1}$ \citep[see][]{Egami2004}. New observations taken
on April 2005 are integrated and standard MIPS procedures were used for the
data reduction. The source extraction is done using DAOPHOT with a limiting
magnitude of 20.3\,mag(AB).

\section{The sample}

\subsection{Sample Construction}
\label{construction}

We have started constructing the multi-colour catalogue of the XMM sources by
cross-correlating the XMM catalogue with the UKIDSS $K$-band catalogue. Since
the vast majority of the XMM sources are AGN, looking in the optical bands
for counterparts to the X-ray sources would introduce a bias against red AGN,
which show increased X-ray to optical ratios but normal X-ray to near-infrared
ratios \citep{Mainieri2002,Brusa2005}. Ideally we should have used the deeper
IRAC $\rm 3.6\,\mu m$ image, however we use the $K$-band due to its smaller
PSF.

\begin{figure}
  \resizebox{\hsize}{!}{\includegraphics{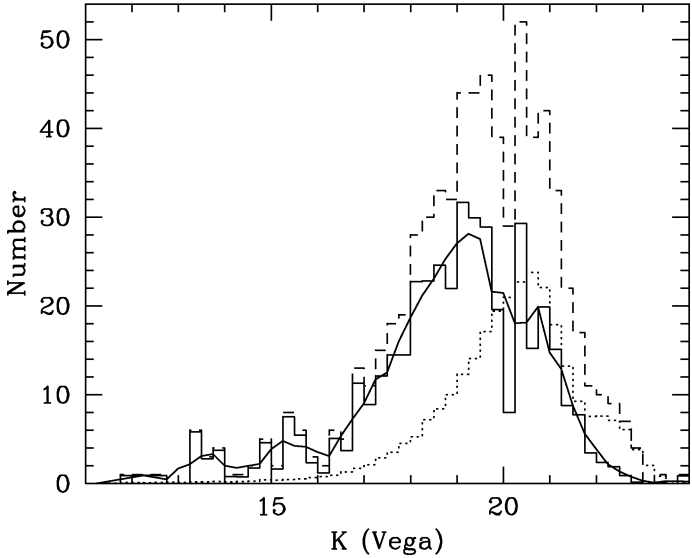}}
  \caption{Various distributions used for the calculation of the likelihood
           ratio of the $K$-band to X-ray counterparts. The histograms
           represent the $total(m)$, $n(m)\pi r_{\rm s}^{2}$, and $real(m)$ with
           the dashed, dotted, and solid lines respectively, while the solid
           line is the smoothed $real(m)$ used for the $LR$ calculation.}
  \label{likelihood5}
\end{figure}

We use the likelihood ratio method \citep{Sutherland1992} to find $K$-band
counterparts to the X-ray sources. The likelihood ratio ($LR$) of a possible $K$
counterpart with magnitude $m$ at a distance $r$ from an X-ray source is
defined as:
\begin{equation}
LR(m,r)=\frac{q(m)f(r)}{n(m)}
\label{LR_eq}
\end{equation}
where $f(r)$ is the combined probability distribution function of the
positional errors of the two catalogues. Here, we assume Gaussian
distributions and define:
\[f(r)=\frac{1}{2\pi\sigma_{1}\sigma_{2}}e^{-\frac{r^{2}}{2\sigma_{1}\sigma_{2}}}\]
where $\sigma_{1}$ and $\sigma_{2}$ are the uncertainties of the positions in
the two catalogues. For the positional uncertainties, we use the published
values of \citet{Brunner2008} for the X-ray sources, a fraction (1/4) of the
FWHM of the source for the optical and IRAC sources, and a constant value of
0.5\arcsec\ for UKIDSS. The distributions used to calculate the likelihood
ratio in Equation\,\ref{LR_eq} are: $n(m)$, the surface density of UKIDSS
objects with magnitude $m(\pm dm/2)$, and $q(m)$ defined as:
\[q(m)=\frac{real(m)}{\sum_{i}real(m)_{i}}Q\]
where
\[Q=\int^{m_{lim}}q(m){\rm d}m\]
is the probability that the counterpart is brighter than the magnitude limit of
the $K$ catalogue (practically the final ratio of X-ray sources with a
$K$ counterpart). $real(m)$ is the expected magnitude distribution of ``real''
counterparts, defined as:
\[real(m)=total(m)-n(m)\pi r_{\rm s}^{2}\]
where $total(m)$ is the total distribution of possible counterparts within the
search radius $r_{\rm s}$. All the above distributions are shown in
Figure\,\ref{likelihood5}.

Given the uncertainties in the positions of the X-ray sources
\citep{Brunner2008} we search for $K$ counterparts with an initial radius of
5\arcsec. The optimum likelihood ratio threshold ($LR_{\rm th}$) we use to
select reliable counterparts is the one which maximises the sum of the mean
reliability and detection rate \citep[see][]{Luo2010}. The reliability $R_i$ of
a possible counterpart $(i)$ is defined as
\[R_i=\frac{LR_i}{\sum LR_j+(1-Q(M_{\rm lim}))}\]
where $j$ refers to the different $K$ counterparts to a specific X-ray source,
and the mean reliability is the mean of the reliabilities of all counterparts
with $LR>LR_{\rm th}$. The detection rate is the ratio of the sum of the
reliabilities of the counterparts with $LR>LR_{\rm th}$ over the number of the
X-ray sources. With $LR_{\rm th}=0.15$, which yields 89.01\% reliability we find
counterparts for 380 X-ray sources. For X-ray sources lacking a $K$-band
counterpart, we repeat the procedure described above using the $R_C$-band
optical catalogue, detecting additional 14 sources, and again for the IRAC and
MIPS catalogues detecting another seven sources.

We finally compile a master catalogue with all the counterparts. We use again
the likelihood ratio method to find the correct optical associations for the
near-infrared counterparts, and vice-versa, this time with an initial search
radius of 3\arcsec. Finally, we visually inspect the optical and infrared
images to spot any obvious mis-identifications and identify any sources which
are missing from the catalogues either because of unreliable photometry (e.g.
saturation) or failure of the source extracting algorithm (e.g. due to a nearby
bright source). We also flag some cases where the optical or infrared
counterpart is a blend of two or more sources.

\begin{table*}
\centering
\caption{Various information of the different catalogues used in this work and
         the correlation of X-ray sources.}
\label{correlations}
\begin{tabular}{ccccc}
\hline\hline
Catalogue                & solid angle & limit              & PSF FWHM & detection rate   \\
                         & deg$^2$     & AB                 & (arcsec) &                  \\
\hline
optical ($R_C$)          & 0.53        & 26.6 ($5\sigma$)   & 0.9     & 385/404 (95.3\%) \\
UKIDSS ($K$)             & 0.78        & 24.0 ($2.5\sigma$) & 0.9     & 382/407 (93.9\%) \\
IRAC ($3.6\,\mu{\rm m}$) & 0.47        & 24.5 ($2\sigma$)   & 2.1     & 381/396 (96.2\%) \\
MIPS ($24\,\mu{\rm m}$)  & 0.84        & 20.3 ($5\sigma$)   & 6.0      & 219/401 (54.6\%) \\
\hline\hline
\end{tabular}
\end{table*}

\begin{figure}
  \resizebox{\hsize}{!}{\includegraphics{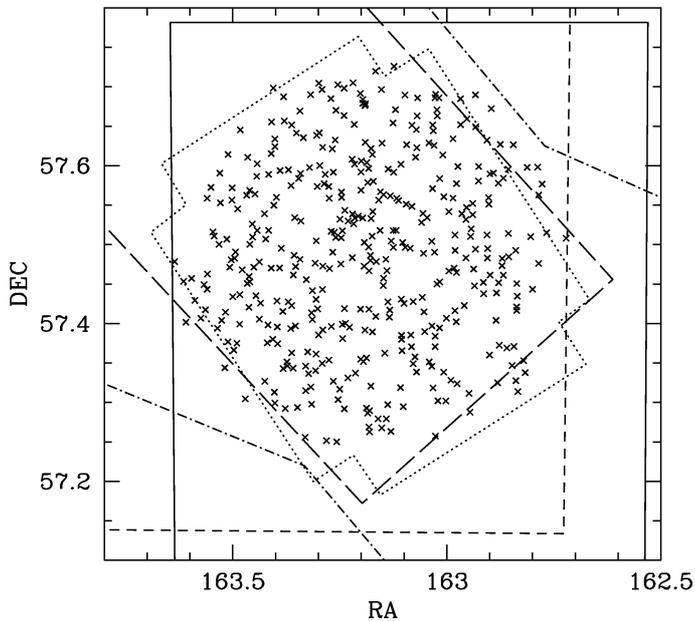}}
  \caption{Coverage of the Lockman Hole region containing the {\it XMM-Newton}
           X-ray sources (plotted in crosses). The dotted, solid, dashed,
           long-dashed, and dot-dashed lines represent the boarders of the
           LBT ($U$-$B$-$V$ bands), Subaru ($R_C$-$I_C$-$z'$ bands), UKIDSS,
           IRAC, and MIPS exposures, respectively.}
  \label{coverage}
\end{figure}

The detection rate of the X-ray sources in the various bands can be seen in
Table\,\ref{correlations}. The highest detection rate is in the IRAC
$3.6\,\mu{\rm m}$ band, and the optical $R_C$-band has a higher detection rate
than the UKIDSS $K$-band. However, the X-rays - $R_C$ associations are made
through the $K$-band, or  better we use $R_C$ counterparts of the $K$-band
sources which in turn are counterparts of the X-ray sources. Differently the
$R_C$ counterparts would not have a high $LR$ in a direct X-ray - optical
counterpart search due to the large number of confusing optical sources. Note
that some X-ray sources are not observed in all available wavelengths. In
Figure \ref{coverage} we can see the distribution of the X-ray sources in the
sky with the areas covered by the different images that map the Lockman Hole
area. We find a reliable optical, near-, or mid-infrared counterpart for
401/409 X-ray sources (98.0\%). Of the 8 sources for which a reliable
counterpart is not found, 2 are associated with diffuse X-ray sources
(XID\,2513, 2514) which mark the positions of galaxy clusters
\citep{Finoguenov2005}, one (XID\,101) is a low-reliability X-ray source
\citep[$L=11$;][]{Brunner2008}, and one (XID\,576) is an off-nuclear source.
The remaining three are not detected in our images.

\subsection{Sample properties}
\label{AGN_select}

\addtocounter{table}{1}

In Table\,\ref{sample_opt} we list the X-ray properties of the 409 XMM sources,
as well as the positions of their optical - infrared counterparts for 401
sources (the position of the optical counterpart in the $R_C$ Subaru image is
given, unless stated otherwise), and some basic optical and infrared
properties. Detailed optical and infrared photometry will be presented in a
future paper together with photometric redshifts, not available at the moment
(Fotopoulou et al. in preparation). For 118 sources spectroscopic redshifts are
available either in literature
\citep{Hasinger1998,Schmidt1998,Lehmann2000,Lehmann2001,Mateos2005} or via a
recent spectroscopic campaign at Keck/DEIMOS (PI: Scoville). For these sources
the X-ray luminosities in Table\,\ref{sample_opt} have been calculated
using the fitted value of $\Gamma$ \citep{Mainieri2002,Mateos2005} or
$\Gamma=2.0$ \citep[see][]{Mainieri2002} when this is not available.

In this paper we want to focus on the observed-frame multi-wavelength
properties of X-ray selected AGN and therefore we make an attempt to
characterise the X-ray sources. In all diagrams we avoid stellar objects,
identified by optical spectroscopy (9 sources with $z=0$ in
Table\,\ref{sample_opt}). One more source (XID\,526) is flagged as a star, based
on its low $R-{\rm m_{3.6\,\mu m}}$ colour \citep[see][]{Ilbert2009}, low X-ray
to optical ratio ($\log(f_{\rm x}/f_{\rm opt})=-1.10$), and soft X-ray spectrum
($HR=-0.73$). Normal galaxies are identified by their low X-ray luminosities
($L_{\rm(0.5-10)\,keV}<10^{42}{\rm erg\,s^{-1}}$) in combination with their soft
X-ray spectra \citep{Bauer2004}. In cases where the X-ray luminosity is not
known they are identified by their low X-ray to optical ratio
\citep{Hornschemeier2003} and their soft spectra. Here, we define a soft X-ray
spectrum as $HR<-0.4$. In order to have a clean AGN sample we flag as galaxies
sources which have $\log(f_{\rm x}/f_{\rm opt})<-1.5$. This value is more
conservative for the AGN selection than the usual limit
($\log(f_{\rm x}/f_{\rm opt})<-2$) and we use it to get a cleaner sample,
knowing that some low luminosity or very heavily obscured AGN may be lost.
Finally, we exclude extended X-ray sources, which are related to galaxy
clusters rather than AGN \citep[11 cases;][]{Finoguenov2005}.

Our final sample consists of 377 X-ray AGN, of these 204 are likely unobscured
and 173 are obscured based on the hardness ratio classification between the
(0.5-2.0)\,keV and (2.0-4.5)\,keV bands and a limit of $HR=-0.4$.

\section{AGN-host contribution}
\label{core_host}

In order to have a measure of the host galaxy contribution to the optical and
infrared flux of the AGN, we use the IRAC colours of the sources of our sample.
The intrinsic mid-infrared spectral energy distribution of a pure AGN
dominated source is a power-law with a spectral index $\alpha\leq-0.5$
\citep{AlonsoHerrero2006,Donley2007} and has characteristic MIR colours.
\citet{Stern2005} define a specific region in the MIR colour-colour diagram
which contains a large fraction (90\%) of optically bright broad-line AGN and
40\% of narrow-line AGN.

\begin{figure*}
  \resizebox{\hsize}{!}{\includegraphics{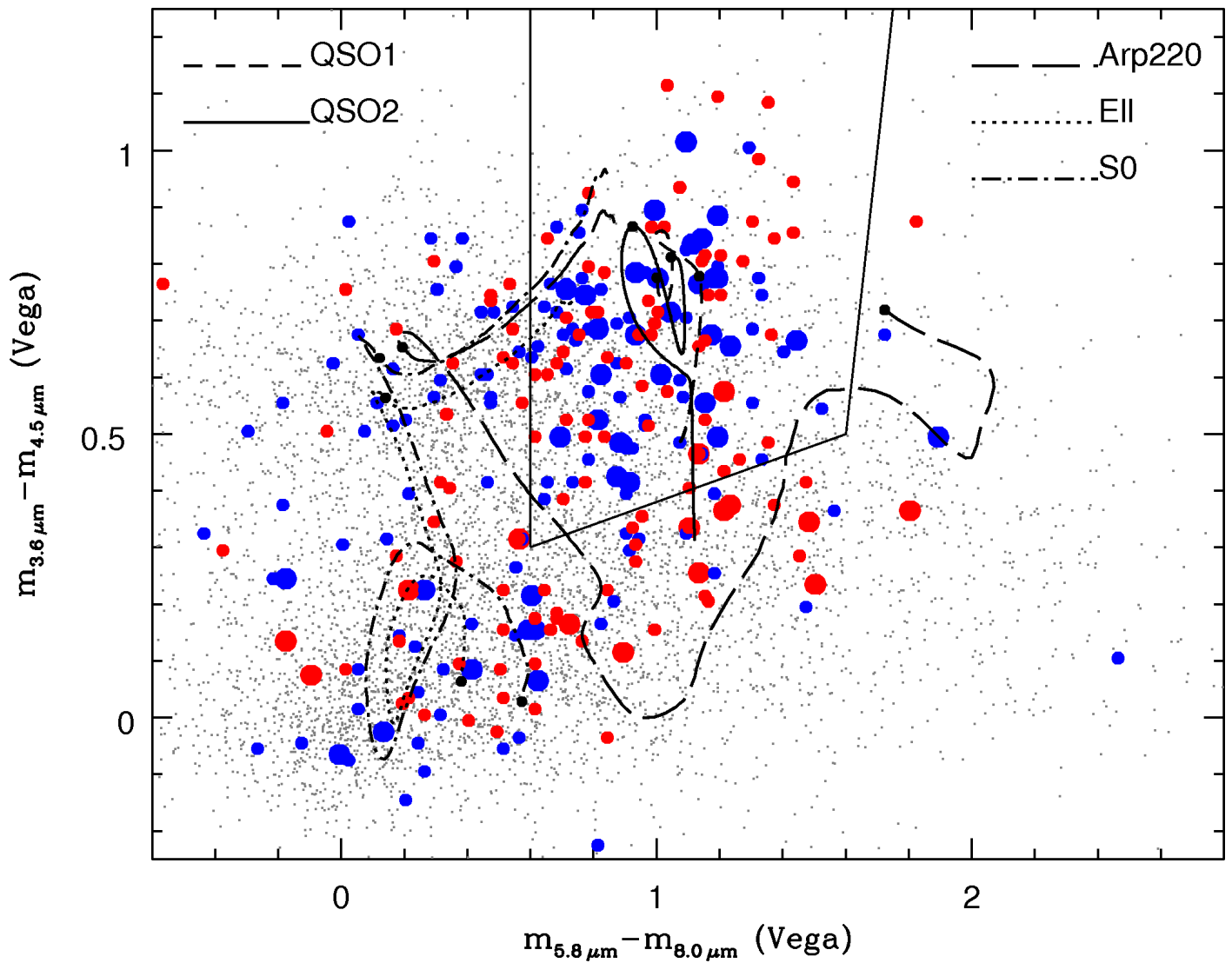}}
  \caption{XMM Lockman Hole sources plotted on the \citet{Stern2005} diagram.
           In blue and red circles are plotted soft and hard X-ray sources
           according to their hardness ratios, while large symbols mark sources
           with bright optical and mid-infrared fluxes ($R_{\rm AB}<21.5$ and
           $f_{\rm 3.6\,\mu m}>12\,\mu{\rm Jy}$) to imitate the selection
           criteria used by \citet{Stern2005}. Grey dots represent the
           mid-infrared colours of all IRAC detections in the Lockman Hole, and
           the lines are tracks of different SED templates with $0<z<4$.}
  \label{stern}
\end{figure*}

In Figure\,\ref{stern} we plot the $[3.6]-[4.5]$ vs. $[5.8]-[8.0]$ colours of
X-ray AGN with IRAC counterparts, marking the region specified by
\citet{Stern2005} with a solid line. Red and blue symbols represent X-ray
obscured and unobscured AGN respectively, while larger circles mark sources
bright both in the optical ($R_{\rm AB}<21.5$) and in $3.6\,\mu{\rm m}$
($f_{\rm 3.6\,\mu m}>12\,\mu{\rm Jy}$) to reproduce the sources used in
\citet{Stern2005}. The lines are
tracks of different spectral energy distributions from $z=0$ to $z=4$ with
black dots marking $z=0$ and $z=2$; the long-dashed line is the SED of Arp\,220,
a dusty starburst galaxy and the prototypical ULIRG, the dotted line is the SED
of an elliptical, the dot-long-dashed line the SED of a S0 spiral, and finally
the short-dashed and solid lines are broad-line QSO1 and QSO2 SEDs respectively.
All SEDs are taken from the SWIRE template
library\footnote{{\tiny http://www.iasf-milano.inaf.it/\~\,polletta/templates/swire\_templates.html}} \citep{Polletta2007}.

We can see in Figure\,\ref{stern} that 277 out of the 377 X-ray selected AGN
are detected in all four IRAC bands, and they are evenly distributed inside and
outside the ``wedge'' which defines typical mid-infrared colours of optically
bright AGN. If we look into their X-ray spectral properties, 79/152 (52.0\%) of
unobscured AGN with full IRAC photometry are inside the wedge and 73/152
(48.0\%) outside. The numbers for obscured objects are 58/125 (46.4\%) and
67/125 (53.6\%) in and out of the mid-IR wedge, respectively. These numbers
change dramatically if we consider only bright sources in the optical and
near-infrared ($R_{\rm AB}<21.5$ and $f_{\rm 3.6\,\mu m}>12\,\mu{\rm Jy}$) to
reproduce the \citet{Stern2005} sample. The fraction of unobscured AGN inside
the wedge rises to 27/37 (73.0\%), while the fraction of obscured AGN inside
the wedge falls to 2/15 (13.3\%). Thus the \citet{Stern2005} criterion is
reliable only in the bright unobscured AGN regime, as was already discussed by
\citet{Barmby2006} in the AEGIS survey, where only 40\% of the X-ray sources
have red ($\alpha<0$) SEDs
\citep[see also][]{Cardamone2008,Brusa2009,Brusa2010}.

In this paper we will use the position in the \citet{Stern2005} diagram not as
a selection criterion for AGN but as an indication of the relative contribution
of the AGN in the optical and infrared colours. We assume that the AGN part of
the SED is always a red power-law \citep[see][]{Elvis1994} being a combination
of (thermal) emission from the heated dust of the circumnuclear torus and
(non-thermal) nuclear emission \citep{Rieke1981}. Its IRAC colours would
reside in the wedge independent from its obscuration for the vast majority of
our AGN. The track of the QSO1 SED (dashed line in Figure\,\ref{stern}) is
always in the wedge, whereas the QSO2 SED (solid line in Figure\,\ref{stern})
is in the wedge for $z<3.87$; only 1/95 of our X-ray AGN with a spectroscopic
redshift have $z>3.87$. On the other hand, the non-AGN tracks are outside the
wedge for $z<2.85$, while only 7/95 (7.3\%) of our X-ray AGN with a
spectroscopic redshift have $z>2.85$. Assuming that there is a link between
X-ray and optical-infrared obscuration, a bright unobscured AGN will have its
infrared colours dominated by the active nucleus, explaining the bright blue
points in Figure\,\ref{stern}. Checking the AGN with spectroscopic redshift
information we see that luminous AGN (with
$L_{\rm x}>10^{44}\,{\rm erg\,s^{-1}}$) with full IRAC photometry are indeed
predominantly unobscured (28/35) objects inside the wedge (32/35). Therefore,
the (optically) bright AGN outside the wedge are low apparent X-ray luminosity
(either obscured or intrinsically faint) AGN at lower redshift ($z<0.8$), where
the host galaxy contribution to the IRAC flux is not negligible both for
obscured and unobscured objects. In the fainter regime, the AGN light is either
intrinsically fainter making the host contribution stronger even for unobscured
systems, or we are at higher redshifts, where the host galaxy luminosity is
increased \citep{Lilly1996}, diluting the observed-frame mid-infrared light of
even brighter unobscured AGN, hence the faint unobscured AGN outside the wedge.
On the other hand, the light of even an obscured AGN could prevail over a weak
host, explaining the faint obscured systems inside the wedge.

\begin{figure}
  \resizebox{\hsize}{!}{\includegraphics{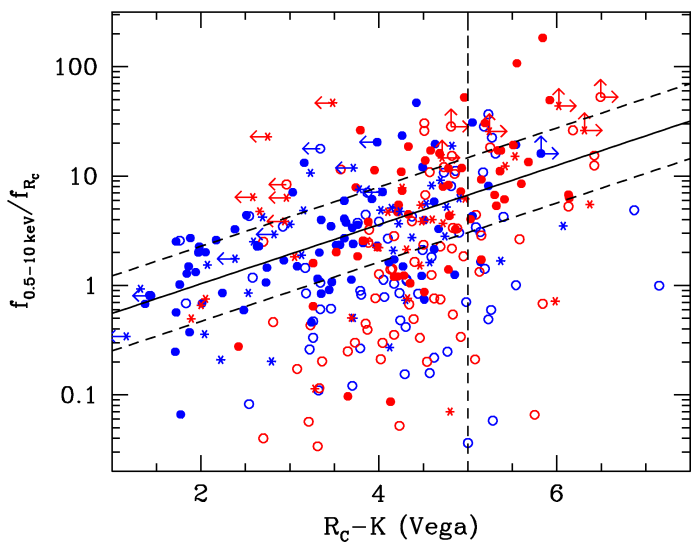}}
  \caption{X-ray to optical flux ratio versus $R_C-K$ colour of the X-ray AGN
           in the Lockman Hole. Blue and red colours represent hard and soft
           X-ray sources respectively according to their hardness ratio and
           different symbols represent the position of the IRAC counterpart in
           the Stern diagram (Figure\,\ref{stern}). Filled circles are sources
           in the wedge (with MIR colours typical of AGN), open circles are
           sources outside the wedge (with MIR colours typical of non-active
           galaxies) and asterisks are sources without full IRAC photometry,
           either because they are fainter than the IRAC limiting magnitude or
           out of the field. The solid line is the best-fit line to the
           positions of the filled circles while the dashed lines mark its
           $1\sigma$ limits.}
  \label{fxfo_RK}
\end{figure}

We further test whether the position of an X-ray source in the IRAC
colour-colour diagram provides a hint about the contribution of the host galaxy
by plotting the X-ray to optical flux ratio against the $R_C-K$ colour in
Figure\,\ref{fxfo_RK}, considering the $R_C-K$ colour to be a proxy of optical
obscuration and extinction. We calculate the X-ray to optical flux ratio using:
\begin{equation}
\log\frac{f_{\rm x}}{f_{R_C}}=\log f_{\rm 0.5-10\,keV}+\frac{R_C({\rm AB})}{2.5}+5.5
\label{fxfo_eq}
\end{equation}
Red and blue circles again mark obscured and unobscured AGN respectively, based
on their X-ray hardness ratio, and filled and open symbols denote the position
of the source in the IRAC colour-colour diagram, with filled symbols
representing sources filling the \citet{Stern2005} criterion. We can see that
filled symbols follow a well defined trend in the $f_{\rm x}/f_{R_C}$ - $R_C-K$
plane. Simultaneously there is a clear dichotomy between obscured and
unobscured objects, obscured being redder in $R_C-K$. We fit a straight line to
the positions of the filled circles of Figure\,\ref{fxfo_RK} using the
orthogonal regression \citep{Isobe1990} as we have uncertainties in both axes
and we get:
\[\log\frac{f_{\rm x}}{f_{R_C}}=(0.27\pm0.05)(R_C-K)-(0.53\pm0.03)\]
which we plot with the solid line, the dashed lines marking the $1\sigma$ area.
The open symbols on the other hand seem to have a more random distribution,
while generally having lower $f_{\rm x}/f_{R_C}$ values for their respective
$R_C-K$. This is a hint that we are detecting a lower AGN contribution in
sources outside the MIR wedge. Also there are very few open symbols with
$R_C-K<3$. This is an area occupied by unobscured QSOs.

\begin{figure}
  \resizebox{\hsize}{!}{\includegraphics{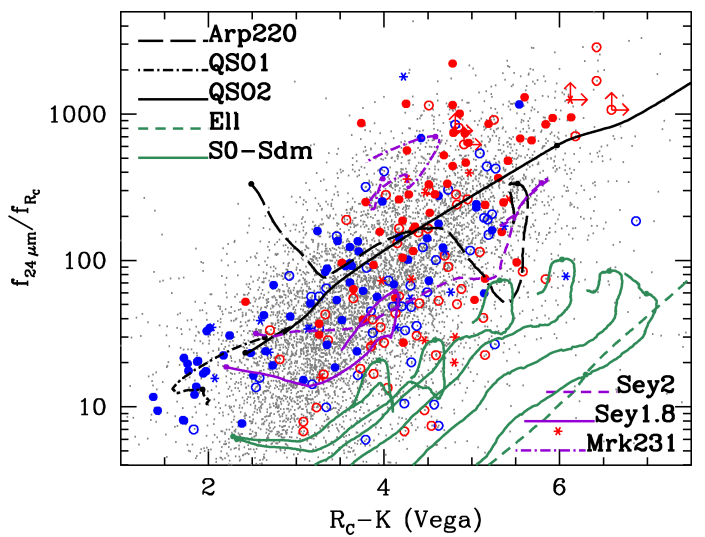}}
  \caption{Mid-infrared ($24\,\mu{\rm m}$) to optical ($R_C$) flux ratio versus
           $R_C-K$ colour of the X-ray AGN in the Lockman Hole. The blue and
           red symbols are identical with Figure\,\ref{fxfo_RK}, while the grey
           dots represent all optical-IRAC-MIPS sources in the Lockman Hole.
           The black lines are the tracks of QSO and Arp\,220 SEDs with
           $0<z<3$, while the violet lines are tracks of SED combinations
           between QSOs and host galaxies (non type-1 Seyferts). Green lines
           from right to left are tracks of elliptical, spirals, and irregular
           galaxies.}
  \label{f24fR_RK}
\end{figure}

The different behaviour of sources inside and outside the IRAC colour wedge is
even more evident when we plot the $24\,\mu{\rm m}$ to optical ($R_C$) flux
ratio against the $R_C-K$ colour in Figure\,\ref{f24fR_RK}. The blue and red
symbols are identical to Figure\,\ref{fxfo_RK}, while with grey dots we plot
the $24\,\mu{\rm m}$ sources with no X-ray detection. The black galaxy tracks
are the same as in  Figure\,\ref{stern} and we also plot tracks of Seyfert-2
and Seyfert-1.8, as well as the Compton-thick AGN Mrk\,231 with dashed,
solid and dot-dashed violet lines respectively. The $24\,\mu{\rm m}$ to optical
flux ratio is often used to detect galaxies with dust obscuration, which
enhances $24\,\mu{\rm m}$ emission while blocking optical light. The most
extreme cases, with $f_{\rm 24\,\mu m}/f_{opt}>1000$ are the ``dust obscured
galaxies'' class \citep[DOGs;][]{Houck2005}, which are usually found in
relative high redshift sources
\citep[$z\simeq 2$;][]{Fiore2008,Dey2008,Pope2008,Fiore2009}.

In Figure\,\ref{f24fR_RK} we can see that the filled and open symbols follow
different distributions. There is again a correlation between
$f_{\rm 24\,\mu m}/f_{R_C}$ and $R_C-K$ for filled symbols, which does not hold
for open symbols; the latter seem to lie on the locus of spirals (moderately
star-forming objects) or optically obscured Seyferts, which may be expected,
assuming that their optical and infrared light is dominated by the host galaxy.
The filled symbols in Figure\,\ref{f24fR_RK} can be fitted with QSO templates,
QSO1 in low $R_C-K$ cases and QSO2 in high $R_C-K$. There is a clear dichotomy
between obscured and unobscured objects both in $R_C-K$ and in
$f_{\rm 24\,\mu m}/f_{R_C}$.

\begin{table}
\centering
\caption{Fractions of obscured (with $HR>-0.4$) X-ray AGN with respect to their
         mid-infrared ($24\mu{\rm m}$) to optical flux ratio and $R_C-K$
         colour. Here an $R_C-K=4$ division is used to denote optically
         obscured and unobscured objects. This is a more modest limit than the
         $R_C-K=5$ used in this paper and elsewhere to select EROs.}
\label{various_fractions}
\begin{tabular}{ccc}
\hline\hline
                                 &     filled     &       open      \\
                                 & (inside wedge) & (outside wedge) \\
\hline
$f_{24\,\mu{\rm m}}/f_{R_C}>100$ & 38/56 (67.9\%) & 18/34 (52.9\%)  \\
$f_{24\,\mu{\rm m}}/f_{R_C}<100$ & 13/57 (22.8\%) & 33/63 (52.4\%)  \\
$R_C-K>4$                        & 40/55 (72.7\%) & 36/64 (56.3\%)  \\
$R_C-K<4$                        & 11/58 (19.0\%) & 15/33 (45.5\%)  \\
\hline\hline
\end{tabular}
\end{table}

In Table\,\ref{various_fractions} we indicate the fraction of obscured objects
of various populations on Figure\,\ref{f24fR_RK} according to their
$f_{\rm 24\,\mu m}/f_{R_C}$ or $R_C-K$ values. All the sources taken into account
are detected in all $R_C$, $K$, IRAC and MIPS bands. We can see that the sources
which comply with the \citet{Stern2005} AGN criteria have high fractions of
obscured objects if they have high $f_{\rm 24\,\mu m}/f_{R_C}$ or $R_C-K$ values
and low fractions otherwise. The correlation of the hardness ratios with
$f_{\rm 24\,\mu m}/f_{R_C}$ or $R_C-K$ is always significant, $>99\%$ according
to K-S tests. On the other hand, sources with IRAC colours not compatible with
AGN do not show any correlation, and we assume that this is due to dilution of
their optical and infrared fluxes by the host galaxy, which smears out
their positions in the $f_{\rm 24\,\mu m}/f_{R_C}$-$R_C-K$ diagram.

The optical-infrared colours of the host dominated AGN cannot be fit with
either QSO or passive galaxy templates (e.g. elliptical or S0). The passive
templates have a very low $f_{\rm 24\,\mu m}/f_{R_C}$ ratio and the QSO very high.
A good fit is a combination of the two; \citet{Pozzi2007} used combinations of
passive galaxy and AGN templates to explain the optical to ${\rm 24\,\mu m}$
SED of a number of obscured quasars
\citep[see also][]{Salvato2009,Merloni2010}. The tracks of the Seyfert-1.8 and
Seyfert-2 templates from \citet{Polletta2007}, which are examples of
moderately luminous AGN with host galaxy contribution explain the high end of
the optical - mid-infrared colours of host dominated AGN. Another possible
alternative is a template with enhanced star formation. The solid green
lines in Figure\,\ref{f24fR_RK} represent the tracks of S0, Sa, Sb, Sc, Sd, and
Sdm templates, from right to left, simulating increasing star formation.
A star-forming template with enhanced dust emission to resemble Arp\,220 can
also explain the most extreme cases, in $R_C-K$ colours, of host-dominated
systems. A combination of a dusty starburst with AGN activity (Mrk\,231)
explains the high-$f_{\rm 24\,\mu m}/f_{R_C}$ sources where both components
contribute to the ${\rm 24\,\mu m}$ flux. It is unclear if the
$24\,\mu{\rm m}$ emission of the host-dominated sources (as well as the
non-X-ray sources in grey dots) comes from AGN or a (dusty) starburst emission;
a measurement of the mid and far-infrared properties to derive its temperature
is needed to clarify this issue.

\section{Discussion}
\label{discussion}

\subsection{X-ray Obscuration of Red AGN}

\begin{figure}
  \resizebox{\hsize}{!}{\includegraphics{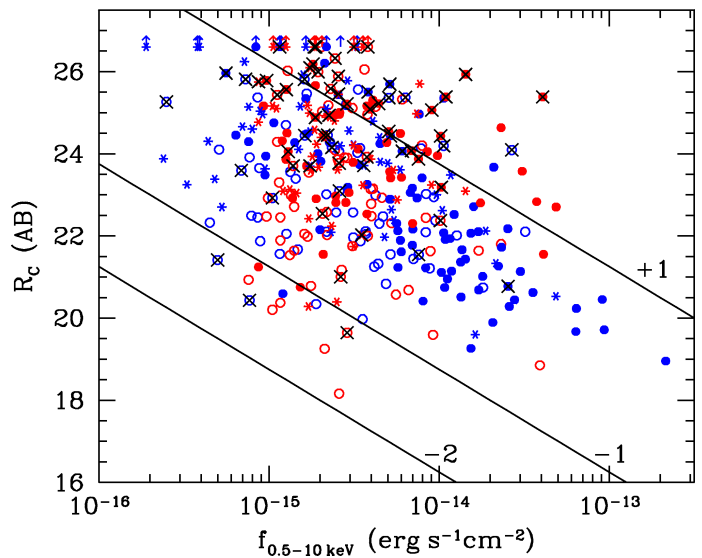}}
  \caption{Optical ($R_C$-band) versus X-ray (0.5-10\,keV) flux of the
           {\it XMM-Newton} sources. With blue and red symbols are plotted soft
           and hard X-ray sources respectively, characterised by their hardness
           ratio, $HR=-0.4$ being the dividing value. Different shapes are used
           according to the positions in the IRAC colour-colour wedge, similar
           to Figures\,\ref{fxfo_RK} and \ref{f24fR_RK}. Crosses mark the
           positions of EROs (with $R_C-K>5$) and lines mark
           $\log(f_{\rm x}/f_{\rm o})=1$, $-1$, and $-2$.}
  \label{fxfo}
\end{figure}

\begin{figure}
  \resizebox{\hsize}{!}{\includegraphics{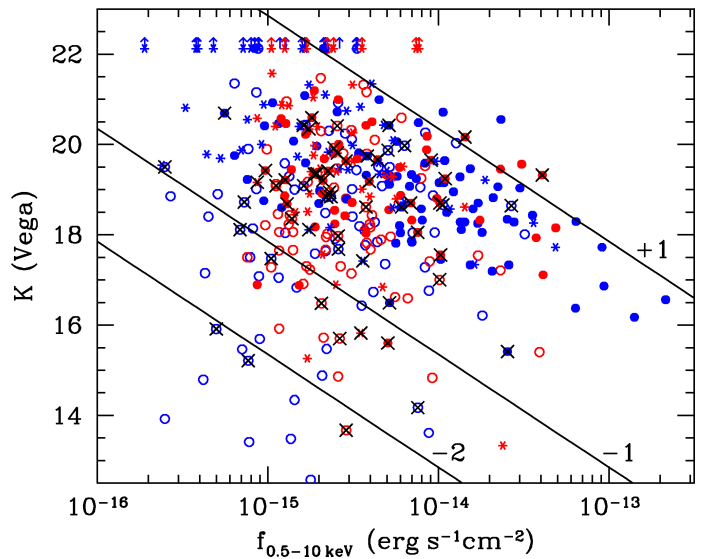}}
  \caption{Same as Figure\,\ref{fxfo} with the UKIDSS $K$-band flux instead of
           $R_C$}
  \label{fxfK}
\end{figure}

In Figures\,\ref{fxfo} and \ref{fxfK} we plot the optical ($R_C$) and
near-infrared ($K$) magnitudes of the X-ray AGN with respect to their
full-band (0.5-10)\,keV fluxes, making a distinction between X-ray obscured and
unobscured objects, as well as their position in the Stern diagram as in
Figures\,\ref{fxfo_RK} and \ref{f24fR_RK}. The lines in Figures\,\ref{fxfo} and
\ref{fxfK} represent
$\log(f_{\rm x}/f_{R_C(K)})=+1,-1,-2$, calculated with Equation\,\ref{fxfo_eq}
and:
\[\log\frac{f_{\rm x}}{f_{K}}=\log f_{\rm 0.5-10\,keV}+\frac{K({\rm Vega})}{2.5}+6.9\]

There are 64 AGN whose optical to near-IR colours ($R_C-K>5$) put them in the
extremely red objects (EROs) regime \citep{Elston1988}, and they are plotted
with a cross in Figures\,\ref{fxfo} and \ref{fxfK}. EROs are examples of
rapidly evolving galaxies, contributing a significant fraction
($\gtrsim 10\%$) of the star formation density of the universe
\citep{Georgakakis2006b}. We can see that they are almost equally distributed
between X-ray obscured and unobscured (39 vs. 25 sources) and tend to have
large X-ray to optical flux ratios but normal X-ray to near-infrared flux
ratios; EROs have $\log(f_{\rm x}/f_{\rm o})=0.7\pm0.7$ and
$\log(f_{\rm x}/f_{K})=-0.2\pm0.7$, while the numbers for the non-ERO AGN
population are $\log(f_{\rm x}/f_{\rm o})=0.2\pm0.6$ and
$\log(f_{\rm x}/f_{K})=0.0\pm0.7$. This is a known property of EROs
\citep{Mainieri2002,Brusa2005} and is suggestive that their red optical to
near-infrared colours are caused by dust extinction which affects the optical
wavelengths but not the near-infrared (or the X-rays in great extent). It is
indicative that \citet{Mainieri2002} do not find any EROs in the bright
XMM - Lockman Hole sample associated with a broad-line AGN; they do on the
other hand find that a third of the ERO sample are unabsorbed in X-rays
($N_{\rm H}<10^{21.5}{\rm cm^{-2}}$). Here, we expand the sample of
\citet{Mainieri2002} by considering multi-wavelength information of the faintest
X-ray sources.

\begin{figure}
  \resizebox{\hsize}{!}{\includegraphics{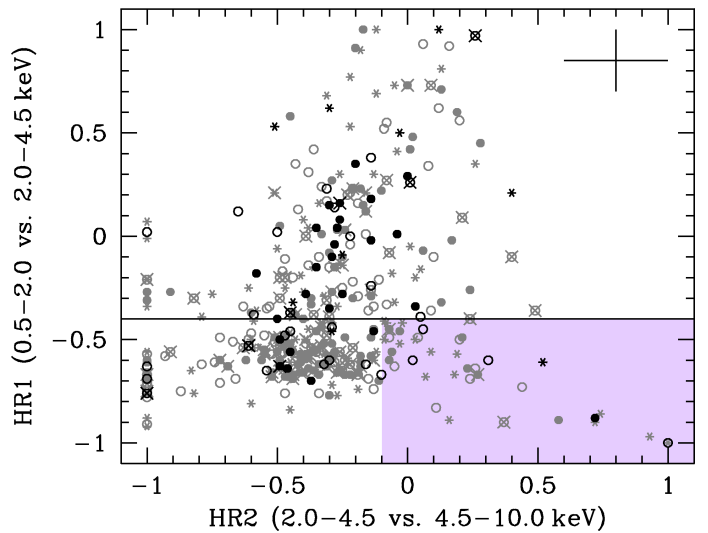}}
  \caption{X-ray colour-colour diagram of the Lockman Hole sources. EROs
           (with $R_C-K>5$) are plotted in black symbols, while the non-ERO
           population in grey. The symbols (open-filled circles and asterisks)
           are identical to Figure\,\ref{fxfo_RK}. The horizontal line at
           $HR1=-0.4$ is the dividing line between soft and hard sources
           adopted for this paper and the purple region is the region where the
           ``leaky absorber'' effect is important (see text). The cross in the
           top right corner shows a typical error-bar with d$HR1=0.15$ and
           d$HR2=0.20$, while crosses on the individual sources mark AGN with
           a known spectroscopic redshift.}
  \label{Xcolcol}
\end{figure}

As we can see in Figures\,\ref{fxfo} and \ref{fxfK} there is a number of EROs
in our sample which are associated with soft X-ray AGN, which reaches 39.0\%
(25/64). We note here that there are two observational effects that might
cause a low hardness ratio of faint sources, without the source really being
soft. The reason for the first effect is that faint sources have a small number
of photons and the fact that {\it XMM-Newton} is more sensitive in the soft
band. This might cause a hard X-ray source to be detected only in the soft
band, even though the harder photons are more abundant. The reason for the
second effect is the nature of photoelectric absorption; when the column
density of the absorbing material is $\rm\lesssim 10^{24}\,cm^{-2}$ then
photoelectric (or Compton-thin) absorption is the dominant mechanism and it
soft X-ray photons are absorbed up to a characteristic energy depending on the
surface density of the obscuring material, causing an absorption turnover in
the X-ray spectrum. In high-redshift (and thus low-flux) objects this turnover
might be redshifted beyond the energy bands where the hardness ratio is
calculated, causing an absorbed AGN to have a low hardness ratio. This redshift
effect becomes important at $z\gtrsim 1.5$ \citep[see][]{Kim2007}. Only three
soft EROs in our sample have a spectroscopic redshift (marked with a cross in
Figure\,\ref{Xcolcol}) and their range is $0.805\leq z\leq 1.018$, while the
typical median redshift for optically selected EROs in deep fields is
$z\simeq 1.2$ \citep{Moustakas2004}.

\begin{figure}
  \resizebox{\hsize}{!}{\includegraphics{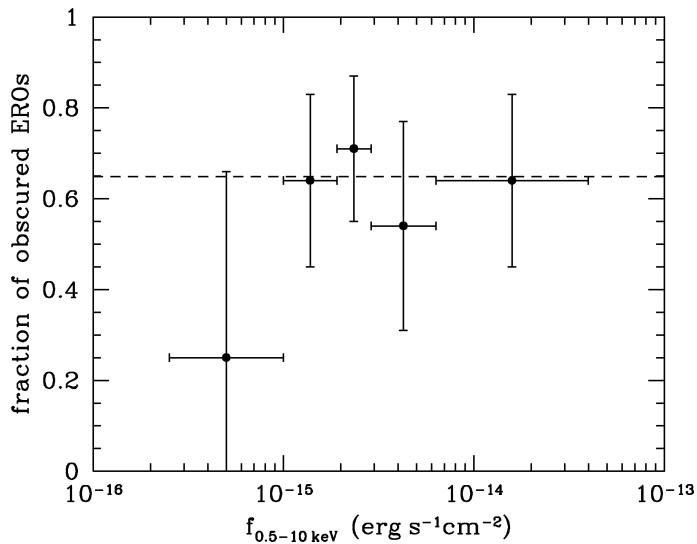}}
  \caption{Variation of the fraction of obscured (with $HR<-0.4$) EROs with
           respect to X-ray flux. The dashed line is at 63.6\%, where the
           overall fraction of obscured EROs with
           $f_{\rm x}>10^{-15}{\rm erg\,s^{-1}cm^{-2}}$ is. We do not see any
           significant dependency of the fraction of obscured EROs with X-ray
           flux, except for sources with
           $f_{\rm x}<10^{-15}{\rm erg\,s^{-1}cm^{-2}}$. The vertical error-bars
           are Poisson estimates, while the horizontal are the bin widths,
           which were chosen to contain the same number of sources.}
  \label{12ratio}
\end{figure}

To further check how both the redshift and the ``low X-ray counts'' effects
affect the number of hard EROs in our sample, we plot it with respect to the
X-ray flux in Figure\,\ref{12ratio}; the dashed line indicates the fraction of
all hard EROs with $f_{\rm x}>10^{-15}\,{\rm erg\,s^{-1}cm^{-2}}$. We do not see
any obvious decrease in the fraction of hard sources with decreasing X-ray flux
which we would expect if the redshift effect was severe (the most distant AGN
are generally expected to have lower X-ray fluxes). We do see it for EROs with
$f_{\rm x}<10^{-15}\,{\rm erg\,s^{-1}cm^{-2}}$, and that is also due to the
higher efficiency of XMM in lower energies and the small number of photons. We
therefore assume that the redshift effect has a limited effect in the fraction
of EROs with $HR<-0.4$, especially in the
$f_{\rm x}>10^{-15}\,{\rm erg\,s^{-1}cm^{-2}}$ regime. If we consider only
sources with $f_{\rm 0.5-10\,keV}>10^{-15}{\rm erg\,s^{-1}cm^{-2}}$, the fraction
of X-ray soft AGN among the EROs is still relatively high (19/56; 33.9\%).

At higher fluxes (lower redshifts), a low hardness ratio of an obscured source
can be caused by a scattered or a thermal component \citep[e.g.][]{Turner1997},
especially since the hardness ratio we use probes relatively low X-ray energies
up to 4.5\,keV. This component is 1-2 orders of magnitude fainter than the
emission from the accretion disk (intrinsic transmission component) and the
absorption turnover would still be recognisable in higher energies.
\citet{Hasinger2007} show that this ``leaky absorber'' spectrum has a higher
``hard'' hardness ratio in the {\it XMM-Newton} bands ($HR2$ between the
2.5-4.5 and 4.5-10.0\,keV bands), with $HR2>-0.1$ when the relative strength
of the scattered component is $<10\%$. In Figure\,\ref{Xcolcol} we plot the
X-ray colour-colour diagram of the X-ray sources
\citep[see also][]{Brunner2008}, using the same symbols as in
Figures\,\ref{fxfo} and \ref{fxfK} and marking EROs with a black colour, the
non-ERO AGN population plotted in grey. We can see that the soft ($HR1<-0.4$)
EROs do not have a different behaviour than the rest of the X-ray AGN. We also
see that there are only six EROs with the ``harder'' hardness ratio being higher
than -0.1, and only two of them (XID\,1566 and XID\,470) have
$f_{\rm 0.5-10\,keV}>10^{-15}{\rm erg\,s^{-1}cm^{-2}}$. We therefore assume that
the ``leaky absorber'' effect does not play an important role for the majority
of the soft EROs. Therefore we assume that they are X-ray unobscured.
Consequently, if the red optical to near-infrared colours of the EROs are due
to dust extinction this indicates that there is a discrepancy between optical
and X-ray absorption in the order of $\sim30\%$. Alternatively the red colours 
could be a result of another process, such as an early-type host or very high
redshift.

In the literature, AGN with red hosts (but not in all cases EROs) are generally
obscured \citep{Rovilos2007,Silverman2008,Georgakakis2008,Pierce2010}, but
there are known cases of unobscured AGN with red hosts \citep*{Georgakakis2006}
where dilution of the AGN optical light by the host galaxy is thought to play an
important role. In Section\,\ref{core_host} we use the mid-infrared colours of
the system to assess the host contribution, and in Figures\,\ref{fxfo} and
\ref{fxfK} we can see that there is a significant number of unobscured EROs
whose optical-infrared colours are AGN dominated. This is more clearly shown
in Figure\,\ref{Xcolcol}, where we can see that while most of the
unobscured EROs (16/25 black objects below the solid line) are host-dominated
and their red colour could come from a red host (either dust-obscured or
evolved; we would need high resolution imaging or optical spectroscopy to
differentiate between the two cases), there is still a non negligible
number\footnote{for 2 sources we do not have complete IRAC photometry} (7/25)
whose red optical-infrared colours are AGN-dominated and a red host could not
explain them.

If we consider red colours as a result of high redshift, their redshift should
be $z>6$ in order for the Lyman break to be redshifted beyond the $R_C$ band.
In such high redshift systems we would expect a high fraction of soft X-ray
sources as a result of the K-correction. In other words, the signature of
obscuration in X-rays, which is a turnover in the X-ray spectrum, would be
redshifted out of our observable window. However, very high redshift objects
are expected to have a high X-ray to optical flux ratio and even though EROs in
general do have high $f_{\rm x}/f_{\rm opt}$, only 8/25 (32.0\%) unobscured EROs
have $f_{\rm 0.5-10\,keV}/f_{R_C}>10$. Moreover, even these 8 cases with
$f_{\rm 0.5-10\,keV}/f_{R_C}>10$ are unlikely to be at $z>6$, as the number of
AGN expected with $z>6$ is much lower; \citet{Barger2003} do not find any $z>6$
AGN in the CDFN \citep[see also][]{Rovilos2010,Aird2010}, while the number
density of $L_{\rm x}=10^{43}-10^{44}{\rm erg\,s^{-1}}$
AGN\footnote{The X-ray luminosity of a $z=6$ source with
$f_{\rm x}=10^{-16}\,{\rm erg\,s^{-1}cm^{-2}}$ and $\Gamma=2$ is
$4\times 10^{43}{\rm erg\,s^{-1}}$} at $z\simeq 6$ is
$\sim10^{-6}{\rm Mpc^{-3}}$ \citep[see also][]{Silverman2008b}. Although
we cannot rule out that there might be some high redshift AGN among the
unobscured EROs (very high redshift AGN are expected to have soft X-ray
spectra, see \citet{Wang2004}) their number would be much lower than 8.

We have seen that light dilution from the host galaxy, as well as redshift
effects can explain the optical - near-IR colours of a number of soft EROs.
There is a number however of sources for which the MIR colours and high X-ray
fluxes argue against such explanations and present a discrepancy between the
optical and X-ray colours. This discrepancy is something not only confined to
the ERO population nor the Lockman Hole survey
\citep{Mateos2005,Szokoly2004,Tozzi2006,Treister2009}. The opposite effect,
X-ray obscured AGN with no signs of optical obscuration has also been reported
\citep{Tajer2007}. In the next section we will try to find an explanation for
the phenomenon considering line of sight effects or a very high dust to gas
ratio.

\subsection{X-ray and optical classification}

\begin{figure}
  \resizebox{\hsize}{!}{\includegraphics{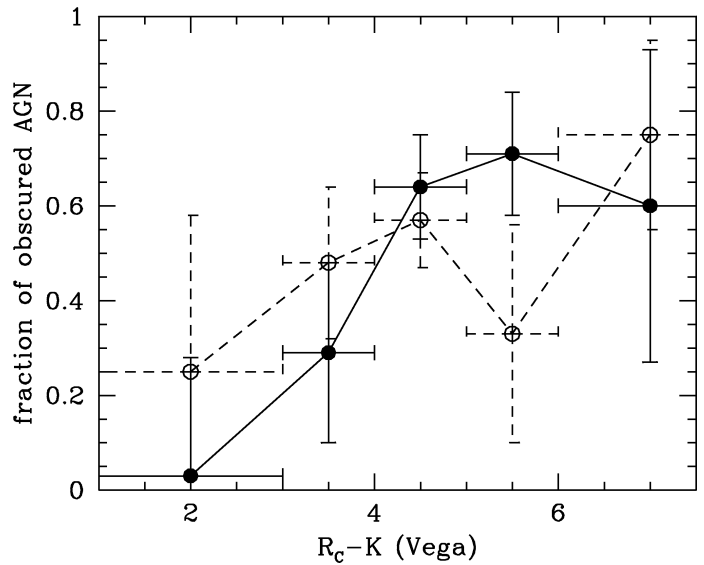}}
  \caption{Fraction of obscured (with $HR>-0.4$) X-ray AGN versus $R_C-K$
           colour. Filled circles and the solid line represent sources which
           mid-infrared IRAC colours satisfy the \citet{Stern2005} criteria and
           are assumed to be AGN-dominated in the optical-infrared, while open
           circles and the dashed line represent sources where the host galaxy
           is assumed to dominate the optical-infrared flux.}
  \label{type_fraction}
\end{figure}

By classifying the AGN using their optical to near-infrared colours we might
be affected by the host galaxy. We use the mid-IR colours to quantify this
by separating sources according to whether they are compatible with AGN SEDs or
not. The fraction of obscured objects is plotted against the $R_C-K$ colour for
sources in (i.e. AGN dominated; filled circle) and out (i.e. host dominated;
empty circle) the Stern wedge in Figure\,\ref{type_fraction}. Clearly the
fraction of obscured systems increases with increasing $R_C-K$, while for
host-dominated systems the fraction seems constant within $1\sigma$.

In host dominated systems we are expecting a loose (if any) correlation between
the optical-infrared colour and the X-ray obscuration properties and this is
what we observe. The colour of the host can come from its dust reddening or
from the colour of its stellar population (see also previous section). A young
stellar population would yield a blue host colour while a red host can come
from both an evolved population and interstellar dust. The interstellar dust
can be the only mechanism that has an effect in the obscuration of the central
engine, hence the loose correlation observed between X-ray obscuration and
star-formation probed by radio \citep{Rovilos2007b}, [OII]-3727
\citep{Silverman2009}, or far-infrared emission \citep{Lutz2010}. Note that
\citet{Rovilos2007b} do find a correlation when considering only X-ray absorbed
cases.

The above assumptions are no longer valid if the $R_C-K$ colours are
characteristic of the AGN rather than the host galaxy. Here, a blue optical
colour means that we see the AGN unaffected by dust, while a red AGN is a
result of its absorption. Indeed, only 1/30 of the bluest AGN (with $R_C-K<3$
and mid-infrared colours in the wedge) is obscured in X-rays (with $HR>-0.4$)
and there is a clear correlation between the optical-to-infrared colour and the
X-ray obscuration. This correlation however does not end up with all red AGN
being obscured, as 27\% (7/26) of AGN with $R_C-K>5$ and mid-infrared colours
in the wedge have $HR<-0.4$. If we take into account only objects with
$f_{\rm x}>10^{-15}{\rm erg\,s^{-1}cm^{-2}}$ to avoid the redshift effect and the
effect of {\it XMM-Newton}'s sensitivity, this ratio becomes 6/24. Moreover,
none of these 6 sources is in the grey area of Figure\,\ref{Xcolcol}, where a
scattered X-ray component is likely to affect the hardness ratio. There is
however one source (XID\,26) close to this area (with $HR2=-0.13$) which shows
signs of X-ray absorption having $\Gamma=1.09\pm0.20$ when a single power-law
fit to the spectrum is performed \citep{Mateos2005}. Discarding this source,
the fraction of X-ray unobscured AGN-dominated sources with
$f_{\rm x}>10^{-15}{\rm erg\,s^{-1}cm^{-2}}$ and $R_C-K>5$ is still above
20\% and they all have $HR2<-0.35$.

This population of reddened but X-ray unobscured AGN can be explained with a
simple orientation-based AGN unification model \citep{Antonucci1993}, since the
region which generates the X-ray emission (accretion disk) is not spatially
coincident with the region where the optical-infrared emission is generated
(inner side of the molecular disk or torus). If we consider a clumpy molecular
disk or torus \citep*[e.g.][]{Elitzur2006,Wada2009}, there can exist a
geometrical set-up where the line-of-sight to the accretion disk is clear,
while obscuring clouds are hindering the direct view to the most prominent
optical - near-IR emitting regions \citep[see Fig.\,5 in][]{Shi2006}, or they
are self-absorbed. \citet{Nenkova2008a} and \citet{Nenkova2008b} have shown
that a clumpy torus can cause a large scatter in the X-ray obscuration
properties of AGN which have similar IR characteristics, due to the fact that
the X-rays come from one physically small region while the IR is the integrated
emission of a large number of clouds in a large area. While the X-ray
absortption is generally higher than what would be expected form the infrared,
the opposite effect for a small fraction of the sources is not ruled out.
Another plausible explanation for these X-ray unobscured red objects is that
interstellar dust obscures the optical flux without affecting the soft X-rays.
The opposite behaviour is reported in the local universe in the form of a very
low $E_{B-V}/N_H$ ratio \citep{Maiolino2001} and explained with large dust
grains \citep{Maiolino2001b}. Here, we have to assume low metallicity dust
which obscures the optical wavelengths but lacks the heavy elements (heavier
than oxygen) which would absorb the X-rays.

\subsection{Properties of AGN host galaxies}

\begin{figure*}
  \resizebox{\hsize}{!}{\includegraphics{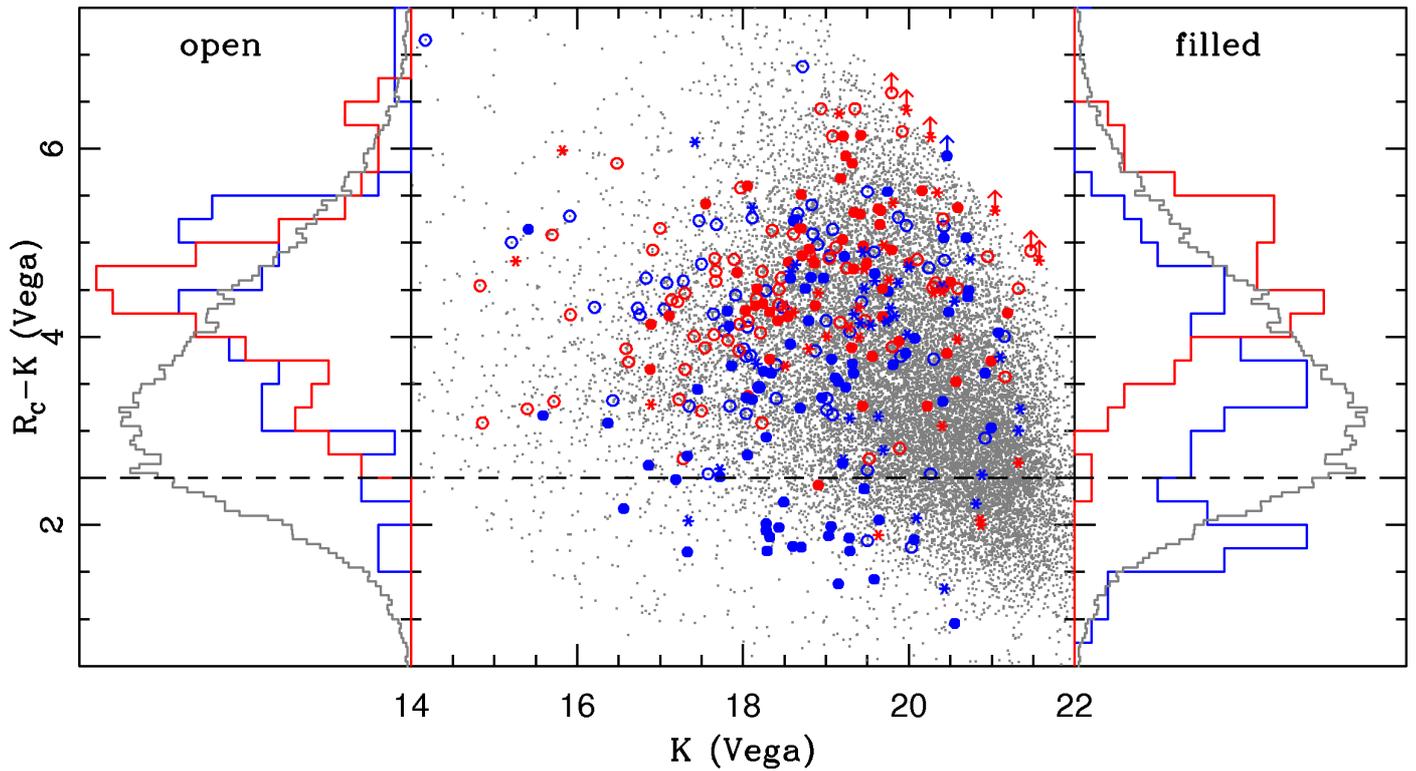}}
  \caption{$R_C-K$ colour versus $K$-band magnitude of X-ray AGN in the Lockman
           Hole. The blue and red symbols are identical to
           Figures\,\ref{fxfo_RK} and \ref{f24fR_RK}, while the grey dots
           represent the colours of optical-UKIDSS normal galaxies. The
           left-side and right-side coloured histograms refer to the open and
           filled symbols, while the grey histogram on both sides refers to the
           grey dots.}
  \label{RK_K}
\end{figure*}

As we can see in Figure\,\ref{f24fR_RK}, the optical to mid-infrared colours of
passive galaxies are quite different, at any redshift, from the typical colours
of AGN-host systems. To check whether the host galaxies of X-ray AGN are a
random sample of sources we plot the $R_C-K$ colour against the $K$ magnitude
in Figure\,\ref{RK_K}. The symbols are identical to Figure\,\ref{f24fR_RK}. We
can see here that the AGN hosts tend to avoid the location of the bulk of the
underlying sources. If we ignore the AGN dominated (filled) unobscured (blue)
sources with $R_C-K<2.5$ (dashed line in Figure\,\ref{RK_K}), which are related
to optical QSOs, the distribution of the AGN hosts is similar to what is found
in the optical rest-frame colour-magnitude diagram
\citep[e.g.][]{Nandra2007,Silverman2008,Georgakakis2008,Hickox2009}, although
there is no clear ``red sequence'' or ``green valley'' because we are using
observed and not rest-frame magnitudes. There is also the trend of redder
objects to be more obscured in X-rays \citep[see][]{Rovilos2007}.

The positions of the AGN hosts in the ``green valley'' of the colour-magnitude
diagram is often linked with the evolutionary sequence of AGN and more
specifically it is argued that it marks the quenching of star formation by
AGN feedback \citep[e.g.][]{Hopkins2005}. However, \citet{Brusa2009} attribute
the red colours to dusty star formation rather than an evolved stellar
population \citep[see also][]{Georgakakis2009}, so that the position in the
colour-magnitude diagram is not indicative of the evolutionary status but more
an indication of the abundance of obscuring material. A closer look in the
histograms of Figure\,\ref{RK_K} reveals that the tendency of redder objects to
be more obscured is confined in the filled symbols (right-side histograms),
indicating optical emission from the AGN, while the colours of the host
galaxies (left-side histograms) show no clear correlation with the obscuration
of the nucleus, something also demonstrated in the two diagrams of
Figure\,\ref{type_fraction}. A K-S test on the $R_C-K$ values of the blue and
red histograms of Figure\,\ref{RK_K}, ignoring the unobscured sources with
$R_C-K<2.5$, gives a null hypothesis probability of
35\% to the ``open'' histograms and 0.009\% to the ``filled''. This means
that the (red) optical colours of obscured AGN are affected by the colour of
the nucleus which gives rise to the correlation
\citep[see also][]{Pierce2010b}, rather than an evolutionary relation between
red hosts and obscured AGN. A weak trend is detectable for host-dominated
systems, which is not statistically significant, it is however still
problematic that a sizeable fraction of red host galaxies host obscured AGN.

We have to note here that a definite answer on the cause of red optical colours
of the host galaxies being dust or an evolved population can be provided by
morphology, which requires high spatial resolution images from space or
adaptive optics, and such observations are not yet available for the Lockman
Hole, or alternatively SED fitting of different templates to disentangle the
different emission mechanisms, which requires complete spectral coverage and
reliable redshifts. A more thorough analysis will take into consideration
spectroscopic (where available) and photometric redshifts and will be presented
in a subsequent paper, when the photometric redshift analysis and the subsequent
SED fitting is completed.

\section{Conclusions}
\label{conclusions}

We have investigated the observed-frame multi-wavelength properties of 377
X-ray selected AGN in the Lockman Hole \citep{Brunner2008}. The optical and
infrared counterparts have been selected from UKIRT, Subaru, and {\it Spitzer}
IRAC and MIPS observations using the likelihood ratio method. Using the
\citet{Stern2005} diagram we made an attempt to assess the relative impact of
the AGN and galaxy emission in the observed optical and infrared flux. Our
results are summarized as follows:
\begin{enumerate}
\item We find an optical and/or infrared counterpart for the vast majority
      (98\%) of the X-ray sources. Of the 8 sources which have no counterpart,
      two are X-ray clusters, one is a low reliability X-ray source, one is an
      off-nuclear source, and three are undetected in our images. The highest
      detection rate is in the mid-infrared ($3.6\,\mu{\rm m}$), where 93.5\% of
      the X-ray sources covered by the IRAC survey are detected.
\item The AGN-dominated and host-dominated sources have different distributions
      in the $f_{\rm x}/f_{R_C}$ vs. $R_C-K$ (Figure\,\ref{fxfo_RK}) and
      $f_{24\,\mu{\rm m}}/f_{R_C}$ vs. $R_C-K$ (Figure\,\ref{f24fR_RK}) diagrams.
      In the latter, the host dominated sources are closer to the distribution
      of non X-ray detected sources, while their colours cannot be explained
      with quiescent galaxy SEDs; either (dusty) star-forming or obscured
      Seyfert SEDs are required.
\item There is a clear correlation between X-ray obscuration and $R_C-K$ colour
      for AGN dominated systems, such that the redder the colors are, the
      higher the fraction of obscured AGN is. However, we don't observe such a
      trend in host-dominated cases (Figure\,\ref{type_fraction}).
\item There is a significant fraction ($\sim1/3$) of EROs which are unobscured
      in X-rays. Moreover, most of them owe their red colours to the AGN and
      not the host galaxy. To explain this population we have to employ
      geometrical arguments or assume special dust properties which absorb the
      optical photons and leave the X-rays intact.
\item The optical - near-infrared colours of X-ray AGN are in general redder
      than those of non X-ray detected objects, and this happens for both AGN-
      and host-dominated systems. We also observe a trend of redder objects
      being more obscured in X-rays, which is more prominent for host-dominated
      systems, indicating the affect of dust obscuration in the
      optical-infrared colours.
\end{enumerate}

\begin{acknowledgements}
G.H. and M.S acknowledge support by the German Deutsche Forschungsgemeinschaft,
DFG Leibniz Prize (FKZ HA 1850/28-1).
We thank M. Brusa for reading the manuscript and providing useful comments.
\end{acknowledgements}

\longtabL{2}{
\begin{landscape}
\begin{longtable}{lccccccccccccl}
\caption{\label{sample_opt} Basic X-ray, optical, and infrared properties of the Lockman Hole X-ray sources}\\
\hline\hline
XID  & $f_{(0.5-10)\,{\rm keV}}$                 & HR & $\log L_{(0.5-10)\,{\rm keV}}$ & RA(optical) & DEC(optical) & $\log\frac{f_{\rm x}}{f_R}$ & $\log\frac{f_{\rm x}}{f_K}$ & $R-K$ & [3.6]-[4.5] & [5.8]-[8.0] & $\log\frac{f_{\rm 24\mu m}}{f_R}$ & $z$ & notes \\
     & ${\rm \times 10^{-15}erg\,s^{-1}cm^{-2}}$ &    & $\rm erg\,s^{-1}$ & deg     & deg      &                                   &                              & Vega  & Vega        & Vega        &                                   &  &       \\
\hline
\endfirsthead
\caption{continued.}\\
\hline\hline
XID  & $f_{(0.5-10)\,{\rm keV}}$                 & HR & $\log L_{(0.5-10)\,{\rm keV}}$ & RA(optical) & DEC(optical) & $\log\frac{f_{\rm x}}{f_R}$ & $\log\frac{f_{\rm x}}{f_K}$ & $R-K$ & [3.6]-[4.5] & [5.8]-[8.0] & $\log\frac{f_{\rm 24\mu m}}{f_R}$ & $z$ & notes \\
     & ${\rm \times 10^{-15}erg\,s^{-1}cm^{-2}}$ &    & $\rm erg\,s^{-1}$ & deg     & deg      &                                   &                              & Vega  & Vega        & Vega        &                                   &  &       \\
\hline
\endhead
\hline
\endfoot
 352 &    2.10 & -0.39 & \ldots & 162.720153 & 57.508367     &   \ldots &   -0.75  &   \ldots & \ldots & \ldots &   \ldots & \ldots     &          \\
 358 &   23.34 & -0.58 &  45.05 & 162.766346 & 57.515112     &    0.56  &    1.49  &    0.95  &  0.56  &  1.08  & $<$0.85  &  2.742$^3$ &          \\
 458 &   23.89 &  0.30 & \ldots & 162.776478 & 57.576667$^1$ &   \ldots &   -1.39  &   \ldots & \ldots & \ldots &   \ldots & \ldots     &          \\
 430 &   36.31 & -0.42 &  44.52 & 162.784495 & 57.562606$^1$ &   \ldots &    0.76  &   \ldots & \ldots & \ldots &   \ldots &  1.540$^3$ &          \\
1522 &    1.89 &  1.00 & \ldots & 162.784408 & 57.476043$^1$ & $>$1.42  &    0.16  & $>$6.41  &  0.37  & \ldots &   \ldots & \ldots     &          \\
 500 &    6.20 & -0.16 & \ldots & 162.787123 & 57.598185$^1$ &   \ldots &    0.50  &   \ldots & \ldots & \ldots &   \ldots & \ldots     &          \\
 217 &   16.01 & -0.56 & \ldots & 162.798959 & 57.442894     &    0.79  &    0.53  &    3.92  &  0.41  &  0.73  &    1.60  & \ldots     &          \\
 357 &    1.88 & -0.28 & \ldots & 162.803303 & 57.510184     &    0.73  &   -0.09  &    5.32  &  0.52  &  0.78  & $<$2.11  & \ldots     &          \\
\end{longtable}
\begin{description}
\item[$^a$] Low quality redshift
\item[$^1$] UKIDSS $K$-band position
\item[$^2$] IRAC $\rm 3.6\,\mu m$ position
\item[$^3$] Redshift from \citet{Schmidt1998}
\item[$^4$] Redshift from \citet{Lehmann2001}
\item[$^5$] Redshift from \citet{Mateos2005}
\item[$^6$] Redshift from Keck/DEIMOS
\item[$^7$] Flagged as star because of low $R-{\rm m_{3.6\,\mu m}}$ colour
\end{description}
\end{landscape}
}

\end{document}